\documentclass[a4paper,fleqn,final]{cas-dc}

\usepackage[T1]{fontenc}
\usepackage[utf8]{inputenc}
\usepackage[main=english]{babel}
\usepackage{textcomp}
\usepackage{hyperref}
\usepackage{amsmath,amssymb}
\usepackage{graphicx}
\usepackage{booktabs}
\usepackage{tabularx}
\usepackage{array}
\newcolumntype{C}{>{\centering\arraybackslash}X}
\usepackage{calc}

\usepackage{threeparttable}
\usepackage{longtable}
\usepackage{adjustbox}
\usepackage[numbers,sort&compress]{natbib}
\usepackage{caption}
\usepackage{etoolbox}
\usepackage{placeins}
\usepackage{needspace}
\usepackage{stfloats}
\usepackage{xurl}
\usepackage{microtype}
\usepackage{lastpage}

\extrafloats{100}
\makeatletter
\g@addto@macro{\UrlBreaks}{\do\_\do\-\do\/\do\.\do\0\do\1\do\2\do\3\do\4\do\5\do\6\do\7\do\8\do\9}
\makeatother

\renewcommand{\ttfamily}{\rmfamily}
\AtBeginDocument{%
  \urlstyle{same}%
}

\usepackage{xcolor}
\definecolor{linknavy}{RGB}{0,70,127}
\hypersetup{
  colorlinks=true,
  linkcolor=black,
  citecolor=linknavy,
  urlcolor=linknavy,
  runcolor=linknavy
}

\AtBeginDocument{%
  \raggedbottom
  \setlength{\abovedisplayskip}{5pt plus 2pt minus 1pt}%
  \setlength{\belowdisplayskip}{5pt plus 2pt minus 1pt}%
  \setlength{\abovedisplayshortskip}{2pt plus 1pt}%
  \setlength{\belowdisplayshortskip}{4pt plus 1pt minus 1pt}%
}

\AtBeginEnvironment{table}{\let\sffamily\rmfamily}
\AtBeginEnvironment{figure}{\let\sffamily\rmfamily}
\AtBeginEnvironment{table*}{\let\sffamily\rmfamily}
\AtBeginEnvironment{figure*}{\let\sffamily\rmfamily}

\ExplSyntaxOn
\RenewDocumentCommand \firstname {}
  { \textcolor{black}{\seq_use:Nn \l_stm_au_seq { ~ }} }

\RenewDocumentCommand \emailauthor { m m }
   {
     \int_gincr:N \g_ead_int
     \seq_gput_right:Nn \g_stm_ead_seq
       {
         { \href{mailto:#1}{\rmfamily #1} }
         \parsename { #2 }
         \space(\eadauthor)
       }
     }

\cs_set:Npn \__first_footerline:
{
  \group_begin:
  \small
  \normalfont
  \ifnum\theblind>0\relax
  \else
  \__short_authors: :~
  \fi
  \itshape Preprint~submitted~to~Frontiers
  \group_end:
}

\RenewDocumentCommand \printorcid { } { }

\cs_set:Npn \__first_head:
{
  \parbox[t]{\textwidth}
  {
    \rule{\textwidth}{0pt}
  }
}

\cs_set:Npn \__cas_head:
{
  \parbox{\textwidth}
  {
    \rule{\textwidth}{0pt}
  }
}

\cs_set:Npn \__cas_foot:
{
  \parbox[t]{\textwidth}
  {
   \rule{\textwidth}{.2pt}\\
   \small
   \normalfont
   \__first_footerline:
   \hfill Page~\thepage {}~of~ \lastpage
  }
}
\ExplSyntaxOff

\makeatletter
\ps@cas
\makeatother

\begin{document}

\makeatletter
\def\bstctlcite{\@ifnextchar[{\@bstctlcite}{\@bstctlcite[@auxout]}}
\def\@bstctlcite[#1]#2{\@bsphack
  \@for\@citeb:=#2\do{%
    \edef\@citeb{\expandafter\@firstofone\@citeb}%
    \if@filesw\immediate\write\csname #1\endcsname{\string\citation{\@citeb}}\fi}%
  \@esphack}
\makeatother
\bstctlcite{IEEEbsTcontrol}

\shortauthors{Farajpoor and Narimani}
\shorttitle{Compact foundation models for floodwater mapping}

\title[mode=title]{Accuracy and efficiency of compact foundation models for surface-water mapping during floods}

\author[1]{Parastoo Farajpoor}
\author[1]{Mohammadreza Narimani}
\cormark[1]
\ead{mnarimani@ucdavis.edu}

\affiliation[1]{organization={Department of Biological and Agricultural Engineering, University of California, Davis},city={Davis},state={CA},postcode={95616},country={USA}}

\cortext[1]{Corresponding author}

\begin{abstract}
Pretrained Earth observation models may reduce the effort needed to map surface water, but their value depends on the alternative against which they are compared. We evaluated Prithvi-EO-2.0-Tiny using the 446 hand-labeled Sentinel-2 chips from 11 flood events in Sen1Floods11. A fine-tuned model, a frozen-encoder model, and an identical randomly initialized architecture were compared with a similarly sized U-Net, a larger Prithvi model, and spectral-index thresholds. Experiments varied the number of training chips, excluded events from model fitting, and measured computational cost and prediction reliability. With 252 training chips, U-Net achieved a water intersection-over-union (IoU) of 0.856 on the official test set, compared with 0.833 for fine-tuned Prithvi-Tiny. Pretraining improved the tiny transformer over random initialization by a reported 0.046 IoU with 25 training chips and 0.023 with 252 chips, but did not reverse the U-Net ranking. Across 11 event-exclusion evaluations, mean IoU was 0.768 for fine-tuned Prithvi-Tiny and 0.777 for U-Net; the paired interval included zero. Fine-tuned Prithvi-Tiny processed a 512$\times$512-pixel chip in 16.1 ms on a GPU and 258 ms on a laptop CPU, compared with 30.6 and 665 ms for U-Net. Errors were concentrated near water boundaries, and high predictive entropy identified many, but not all, erroneous pixels. These results support a specific role for compact pretrained models: lower measured computational demand with competitive accuracy, rather than consistent superiority. Event-dependent performance, shared hyperparameter selection, and the distinction between observed water and newly inundated land delimit the conclusions.
\end{abstract}


\maketitle

\section{Introduction}\label{introduction}

Satellite maps made during floods help identify where water is present when field observations are incomplete or difficult to obtain. Their value is not limited to drawing an outline: they inform the interpretation of exposure, the prioritization of further observation, and the assessment of hydrological models. Global studies have documented substantial populations exposed to flooding and an increase in the population observed within large flood extents \citep{tellman2021satellite,rentschler2022flood}. For such applications, a useful mapping method must be accurate enough for its purpose, feasible to run, and explicit about what the imagery cannot reveal.

Optical images provide a relatively direct view of exposed surface water. Water commonly has low near-infrared and shortwave-infrared reflectance, which motivates the normalized difference water index (NDWI) and modified NDWI (MNDWI) \citep{mcfeeters1996ndwi,xu2006mndwi}. These indices are inexpensive and interpretable, but their thresholds are sensitive to scene composition and to surfaces that resemble water spectrally. Learned classifiers can use richer spectral and spatial information \citep{isikdogan2017surface}. Neither approach removes the physical limitations of optical observations: clouds and vegetation can conceal inundation, and water-land mixtures complicate classification at boundaries \citep{shastry2023mapping}. Accordingly, a map of water visible in an image is not automatically a map of all flooding on the ground.

Deep-learning flood mapping has developed around encoder-decoder networks and increasingly diverse benchmark datasets \citep{bentivoglio2022deep}. Sen1Floods11 provides paired observations and reference masks from multiple flood events, including a geographically separate Bolivia evaluation set \citep{bonafilia2020sen1floods11}. Optical flood-segmentation studies have demonstrated the usefulness of this approach, while cross-dataset assessments show that success on one benchmark does not settle transferability to another \citep{portalesjulia2023global,portalesjulia2025understanding}. A strong within-dataset score can coexist with failures in landscapes, water conditions, or labeling conventions that differ from the training examples.

This concern is not specific to floods. The systematic review of Sentinel-2 crop-yield estimation by \citet{narimani2026yield} also organizes evidence around heterogeneous targets, input representations, and validation settings. The broader lesson is that a model's usefulness must be judged against the target and evaluation design actually being studied.

Geospatial foundation models offer a different starting point for learning. Their encoders are pretrained on large image collections, then adapted to a smaller labeled task. Prithvi-EO-2.0 applies this approach to multispectral, multitemporal Earth observation imagery, and TerraTorch provides tools for downstream adaptation \citep{szwarcman2024prithvi,gomes2025terratorch}. The attraction is practical: a pretrained representation may make better use of a small reference set. However, this proposition contains two separate questions. Does pretraining improve the same network relative to random initialization, and does that pretrained network outperform a strong task-specific alternative? These questions require different controls.

A further distinction is whether the encoder itself is run and adapted or whether a classifier uses representations computed in advance. For example, \citet{narimani2026alphaearth} used AlphaEarth embeddings for crop-system mapping. Here, the Prithvi network operates directly on the supplied multispectral images; frozen and fine-tuned encoders are therefore explicit experimental conditions rather than interchangeable descriptions of an embedding dataset.

Recent evidence argues against treating foundation-model superiority as a given. PANGAEA found that geospatial foundation models do not consistently outperform supervised baselines across tasks and data regimes \citep{marsocci2026pangaea}. Flood-specific work has examined sensor-dependent foundation-model performance and hybrid designs that combine pretrained features with convolutional detail \citep{kaushik2026assessing,kostejn2025uprithvi}. These studies motivate careful comparisons rather than a claim that another application of Prithvi is inherently novel. For a compact model, the relevant contribution is a controlled account of what is gained, what is lost, and under which evaluation conditions.

Geography is central to that account. Nearby samples often share landscape structure and acquisition conditions, so spatially dependent validation can make an environmental mapping method appear more transferable than it is \citep{roberts2017crossvalidation,ploton2020spatial}. The distinction also matters in hazard applications outside flooding: a spatially validated analysis of wildfire structure loss reported a substantial change in apparent performance when evaluation moved beyond random partitions \citep{farajpoor2026palisades}. Event exclusion is useful for exposing such differences, but its interpretation must include the origin of the hyperparameters as well as the images used for gradient updates.

This study evaluates the compact Prithvi-EO-2.0-Tiny encoder with a segmentation decoder against a similarly sized U-Net, spectral-index baselines, an identical randomly initialized transformer, and a larger Prithvi reference model. We ask whether pretraining improves the compact architecture, whether that improvement translates into an advantage over U-Net with fewer fine-tuning labels, and how accuracy and reliability change when an event is excluded from fitting. We also measure inference on a GPU and a laptop CPU. The central aim is to separate the benefit of pretraining from the choice of architecture and from the cost of running the resulting model.

\section{Materials and methods}\label{materials-and-methods}

\begin{figure*}[!t]
\centering
\includegraphics[width=\textwidth]{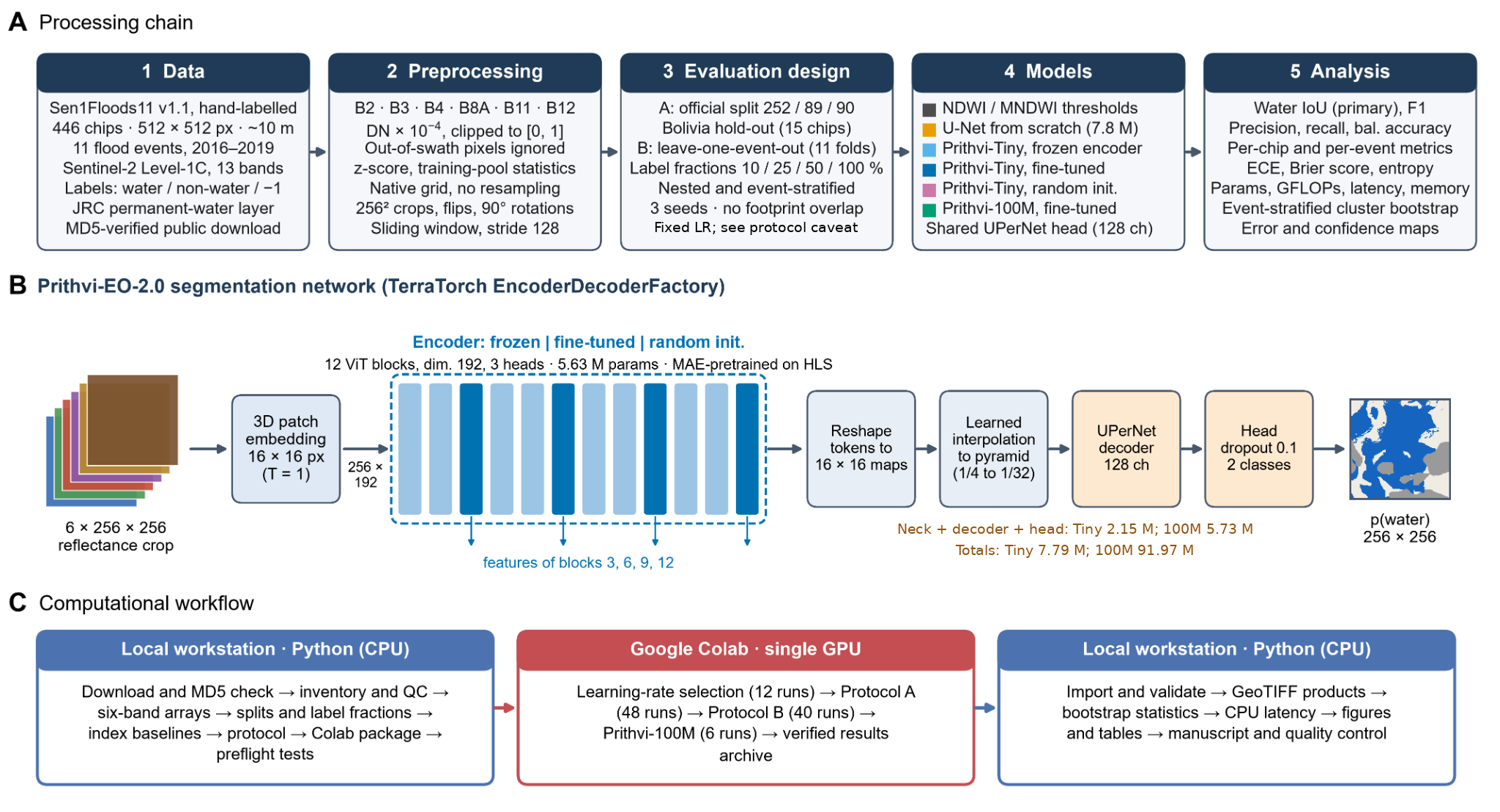}
\caption{Processing, architectures, and computational workflow. (A) Data preparation, evaluation allocations, methods, and diagnostics. The leave-one-event-out shorthand refers to exclusion from the fixed benchmark training/validation pools; global learning rates were selected once under Protocol A and reused. (B) Single-date six-band Prithvi segmentation. Tiny features are taken from blocks 3, 6, 9, and 12 and decoded through a feature pyramid and UPerNet head. The decoder family is shared, but the non-encoder parameter count differs: approximately 2.15 M for Tiny and 5.73 M for the larger model. (C) Local preparation, Colab training, and local analysis. The 106 reported training runs comprise 12 selection runs and 94 evaluation runs.}
\label{fig:methodology}
\end{figure*}

Figure~\ref{fig:study-area} locates the study events, Figure~\ref{fig:methodology} summarizes the workflow, and Figure~\ref{fig:dataset} describes the imagery and reference labels. The two evaluation protocols distinguish unseen chips from events excluded from model fitting.

\subsection{Flood-event imagery and reference labels}\label{flood-event-imagery-and-reference-labels}

\begin{figure*}[!t]
\centering
\includegraphics[width=\textwidth]{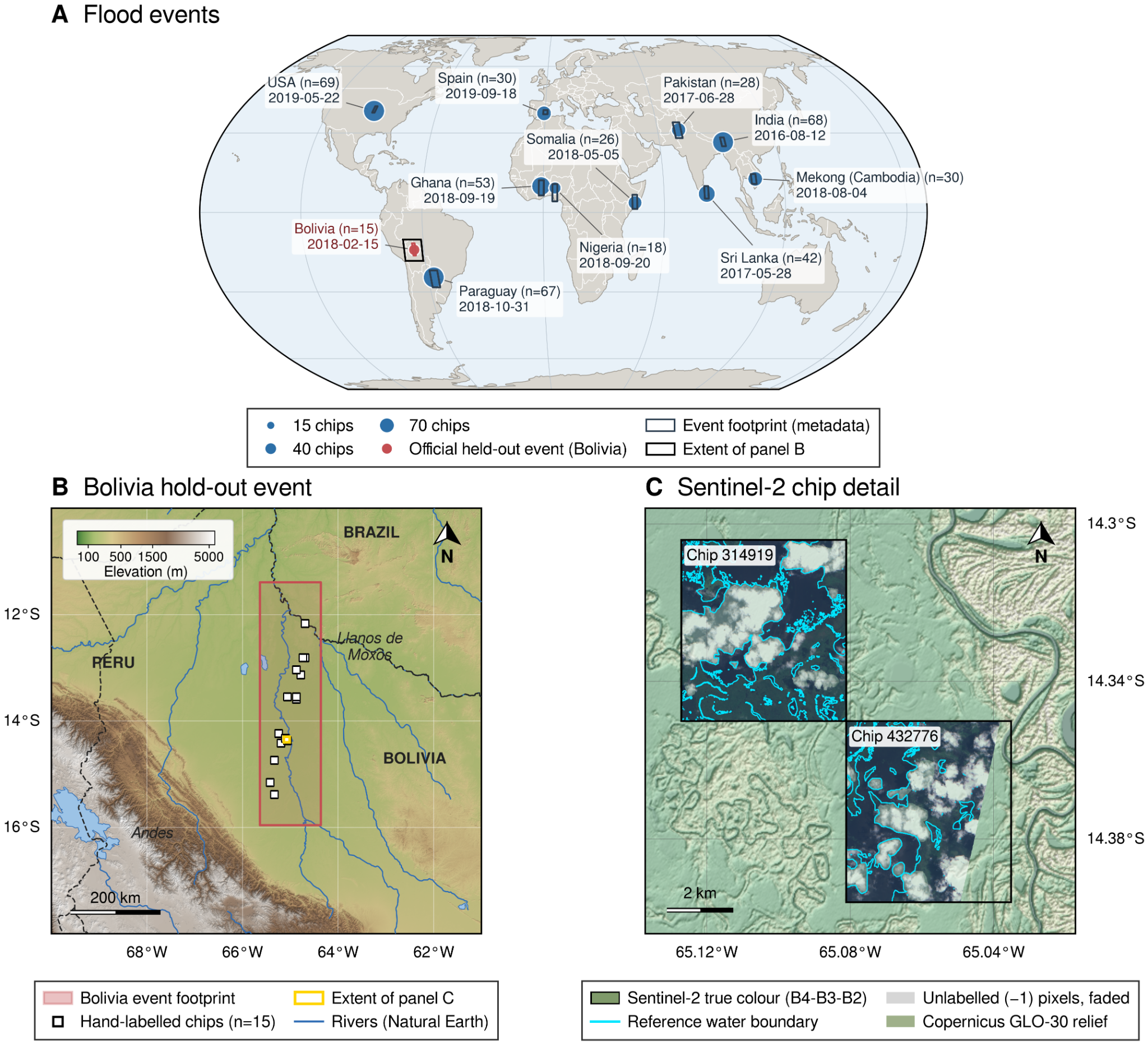}
\caption{Geographic coverage and the Bolivia holdout. (A) Locations, image dates, and hand-labeled chip counts for the 11 Sen1Floods11 events; symbol area scales with chip count. Bolivia is the Protocol A event holdout. (B) Bolivia chip locations and the metadata event footprint over Copernicus GLO-90 relief, with Natural Earth boundaries and rivers. (C) Two example chips, their reference-water boundaries, and Copernicus GLO-30 relief; unlabeled pixels are faded. Panel A uses an Equal Earth projection; the regional panels show WGS 84 geographic coordinates. The source cartography uses vertical exaggeration of 3 for regional shading and 12 for local shading, not for quantitative terrain analysis. Footprints describe the supplied event/chip coverage, not independently measured flood extents.}
\label{fig:study-area}
\end{figure*}

The study used the hand-labeled Sentinel-2 subset of Sen1Floods11 version 1.1 \citep{bonafilia2020sen1floods11}. It contains 446 image chips, each 512$\times$512 pixels, from 11 events observed between 2016 and 2019. Figure~\ref{fig:study-area} shows the locations and the Bolivia evaluation region; Table~\ref{tab:events} gives the acquisition dates and sample allocations. The source images contain 13 Sentinel-2 Level-1C bands, stored as integer top-of-atmosphere reflectance scaled by 10,000. They use a geographic WGS 84 grid with a pixel spacing of approximately $8.98\times10^{-5}$ degrees. This is about 10 m, but the east-west ground distance varies with latitude; no fixed square-kilometer throughput is inferred from chip counts.

\begin{table*}[!t]
\centering
\caption{Event composition and the two evaluation allocations.}
\label{tab:events}
{\small\setlength{\tabcolsep}{3.5pt}\renewcommand{\arraystretch}{1.05}
\begin{tabularx}{\textwidth}{@{}lCCCCCCC@{}}
\toprule
Event & S2 date & Chips & A: train / valid / test & B: train / valid & Water (\%) & Unlabeled (\%) & Permanent water (\%) \\
\midrule
Ghana & 2018-09-19 & 53 & 31 / 11 / 11 & 221 / 78 & 6.3 & 33.8 & 1.4 \\
India & 2016-08-12 & 68 & 40 / 14 / 14 & 212 / 75 & 12.4 & 16 & 2.7 \\
Cambodia & 2018-08-04 & 30 & 18 / 6 / 6 & 234 / 83 & 24.1 & 9.3 & 7.6 \\
Nigeria & 2018-09-20 & 18 & 10 / 4 / 4 & 242 / 85 & 23.5 & 14.3 & 1.9 \\
Pakistan & 2017-06-28 & 28 & 16 / 6 / 6 & 236 / 83 & 7.5 & 26.7 & 2.5 \\
Paraguay & 2018-10-31 & 67 & 39 / 14 / 14 & 213 / 75 & 8.4 & 5 & 2.9 \\
Somalia & 2018-05-05 & 26 & 15 / 5 / 6 & 237 / 84 & 6.5 & 21.1 & 0 \\
Spain & 2019-09-18 & 30 & 18 / 6 / 6 & 234 / 83 & 13.2 & 1.7 & 6.3 \\
Sri Lanka & 2017-05-28 & 42 & 24 / 9 / 9 & 228 / 80 & 11.8 & 9.5 & 7.9 \\
USA & 2019-05-22 & 69 & 41 / 14 / 14 & 211 / 75 & 4.6 & 2.4 & 1.5 \\
Bolivia & 2018-02-15 & 15 & hold-out & 252 / 89 & 15.9 & 27.1 & 1.5 \\
Total & 2016-2019 & 446 & 252 / 89 / 90 & Varies & 10.60 & 13.63 & --- \\
\bottomrule
\end{tabularx}}\\[2pt]
\parbox{\textwidth}{\footnotesize A: official benchmark pools, with 15 additional Bolivia holdout chips. B: counts after excluding the target event from the fixed A training/validation pools; Bolivia does not enter those pools. For non-Bolivia targets, nine other non-Bolivia events remain. Water and permanent-water percentages refer to labeled pixels; unlabeled percentages refer to all pixels. Event percentages are the supplied event summaries; the total row is the supplied pooled summary, not an average of event percentages. Permanent-water membership is from the distributed JRC-derived layer.}
\end{table*}

Reference values are 1 for water, 0 for non-water, and -1 for unlabeled pixels. The water class includes permanent water and water observed during the flood event. It does not identify newly inundated land by itself. A co-registered permanent-water layer distributed with the benchmark and derived from the JRC Global Surface Water product was used only to stratify errors \citep{pekel2016highresolution}. It was not a predictor. Water outside this permanent-water layer is called other reference water below, because absence from that layer is not independent proof of flood-induced change.

\begin{figure*}[!t]
\centering
\includegraphics[width=\textwidth]{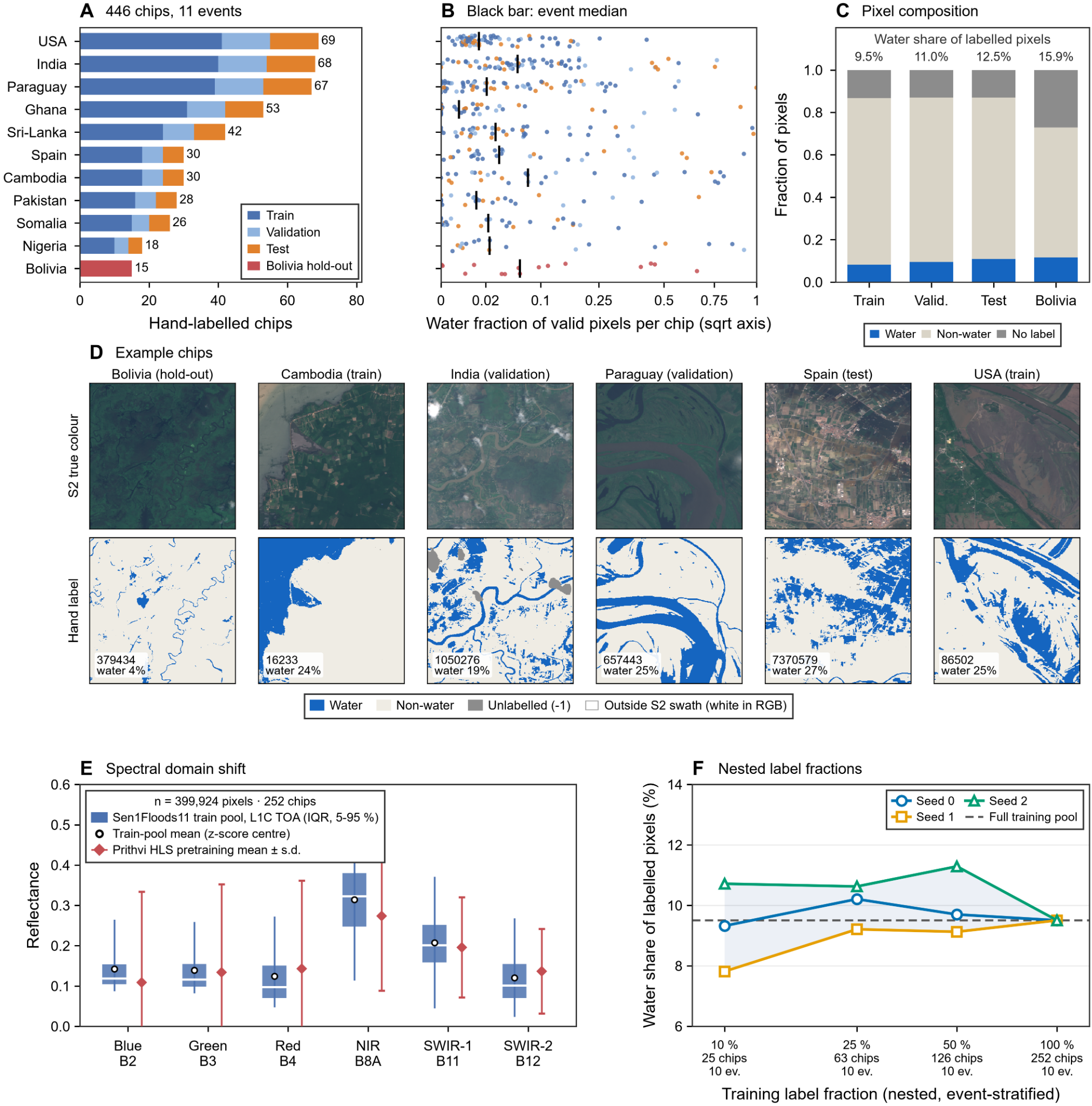}
\caption{Dataset composition and input distributions. (A) Chip counts by event and Protocol A partition. (B) Valid-water fraction per chip, with event medians and a square-root horizontal scale; chips without valid labels are omitted. (C) Pixel composition by split; annotations give water as a fraction of labeled pixels. (D) Selected true-color images and hand labels, including the distinct treatment of unlabeled and out-of-swath areas. (E) Training-pool reflectance distributions compared with the reported HLS pretraining statistics; unequal distribution summaries are descriptive, not a formal domain-shift test. (F) Water fractions of the nested event-stratified training subsets. All training events remain represented, but both subset composition and optimization vary with seed.}
\label{fig:dataset}
\end{figure*}

The inventory recorded aligned image-mask grids, five chips without valid labels, and 52 chips without reference water. Across the collection, 13.63\% of pixels are unlabeled and 10.60\% of labeled pixels are water. In 30 chips, part of the image lies outside the Sentinel-2 swath. The 60,454 labeled pixels with no signal in any band in these areas were excluded for every method, rather than treated as optically observable land or water. Figure~\ref{fig:dataset} summarizes the resulting label distribution. Label provenance is important when comparing hazard benchmarks: the active-fire dataset of \citet{mitra2026firedata}, for example, uses spectral-rule-derived labels, whereas the present analysis uses the benchmark's hand-labeled water masks. The two forms of supervision should not be treated as equivalent evidence.

\subsection{Input representation and preprocessing}\label{input-representation-and-preprocessing}

The six input bands were blue B2, green B3, red B4, narrow near-infrared B8A, and shortwave-infrared B11 and B12. B8A, rather than the wider B8 band, matches the intended Prithvi spectral correspondence. Digital numbers were clipped to [0, 10,000] and multiplied by $10^{-4}$. Images and masks remained on the distributed grid without further resampling. For networks, each band was standardized using the image mean and standard deviation of the relevant training pool; out-of-swath pixels were set to zero after standardization.

This preprocessing does not make the inputs identical to Prithvi's pretraining data. Harmonized Landsat-Sentinel (HLS) provides surface reflectance on a 30-m grid \citep{claverie2018harmonized}, whereas these inputs are Level-1C top-of-atmosphere observations on the supplied finer grid. In Figure~\ref{fig:dataset}E, training-pool band means differ from the HLS pretraining means by -0.019 to +0.041 reflectance, and their standard deviations are 1.3-3.1 times smaller. These descriptive differences indicate an input-domain mismatch; they do not isolate its effect on accuracy. Normalization used the full training-image pool even in reduced-label experiments, so those experiments reduce labeled fine-tuning chips rather than the amount of unlabeled imagery available.

\subsection{Evaluation design and label fractions}\label{evaluation-design-and-label-fractions}

Protocol A preserved the benchmark allocation: 252 training chips, 89 validation chips, and 90 test chips from ten events, plus 15 Bolivia chips as a separate event-level evaluation set. The official test chips therefore represent events seen during training, even though the chips themselves were not used to fit the models. Bolivia was absent from both fitting and validation. We use seen-event test and Bolivia consistently to distinguish these settings.

Protocol B excluded each target event from model fitting and evaluated all its available chips. The reported fold sizes in Table~\ref{tab:events} show how this was implemented: the target event's chips were removed from the fixed Protocol A training and validation pools. Bolivia remained outside those pools throughout. Thus, non-Bolivia target folds used nine other non-Bolivia events, not all ten other events in the complete dataset. The Bolivia evaluation reused the Protocol A seed-0 models. We retain the shorthand leave-one-event-out (LOEO) for correspondence with the figures, but interpret it as event-excluded fitting within fixed benchmark pools.

There is a further limitation to geographic independence. Learning rates were selected once using the full Protocol A validation pool and then reused in Protocol B. For non-Bolivia folds, this earlier selection could include validation observations from the subsequently excluded event. Protocol B therefore tests exclusion from gradient fitting and fold-specific checkpoint selection, but it is not fully nested, event-blind hyperparameter selection. Bolivia remains the cleaner event-level holdout with respect to this shared tuning step. This distinction is important when interpreting geographic transfer rather than merely describing nonoverlapping training files.

For label efficiency, nested subsets contained 25, 63, 126, and 252 training chips, approximately 10\%, 25\%, 50\%, and 100\% of the Protocol A pool. Whole chips were sampled with event stratification, retaining all ten training events. Seeds 0, 1, and 2 controlled subset sampling and network training. Validation and evaluation sets were fixed. Consequently, variation across seeds combines the effects of sample composition and optimization; it does not identify either component separately. The 89 validation labels and the full-label learning-rate selection were retained at every fraction. The curves therefore quantify fine-tuning-label sensitivity under shared model selection, not a complete annotation-budget comparison.

\subsection{Spectral and neural comparison methods}\label{spectral-and-neural-comparison-methods}

NDWI and MNDWI were computed as in Equation~\eqref{eq:ndwi}, using B8A as the near-infrared input for consistency with the six-channel networks \citep{mcfeeters1996ndwi,xu2006mndwi}.

\begin{equation}
\text{NDWI}=\frac{\rho_G-\rho_N}{\rho_G+\rho_N},\qquad \text{MNDWI}=\frac{\rho_G-\rho_S}{\rho_G+\rho_S}
\label{eq:ndwi}
\end{equation}

Here, $\rho_G$ is green-band reflectance (B3), $\rho_N$ is narrow near-infrared reflectance (B8A), and $\rho_S$ is shortwave-infrared reflectance (B11). Each index was thresholded to label water. Three alternatives were evaluated: the fixed threshold zero; one threshold fitted on training labels on a grid with step 0.005 to maximize aggregated water IoU; and an unsupervised threshold estimated separately from each chip's in-swath histogram using Otsu's method \citep{otsu1979threshold}. Index fitting used the same labeled subsets as the corresponding neural experiments and no test labels.

The convolutional baseline was a six-channel U-Net trained from scratch \citep{ronneberger2015unet}. It used four downsampling stages, base width 32, two $3\times 3$ convolutions with batch normalization per stage, transposed-convolution upsampling, and skip connections. Its 7.76 million parameters make it a useful size-matched comparator for the complete Tiny segmentation network, rather than for the encoder alone.

Prithvi-EO-2.0-Tiny was loaded through TerraTorch with the public pretrained weights \citep{szwarcman2024prithvi,gomes2025terratorch}. Its encoder has 12 transformer blocks, embedding dimension 192, and approximately 5.63 million parameters. With one time step and $16\times 16$-pixel patch embeddings, a 256$\times$256 crop produces 256 spatial tokens. Optional time and location metadata were not supplied. Features from blocks 3, 6, 9, and 12 were reshaped into spatial maps, converted into a feature pyramid, and passed to a 128-channel UPerNet decoder and two-class head with dropout 0.1 \citep{xiao2018unified}. This is actual network inference and adaptation, not classification of precomputed embeddings. The masked-autoencoder pretraining principle is described by \citet{he2022masked}.

Three Tiny variants shared the architecture. The frozen variant updated only the neck, decoder, and head; the fine-tuned variant updated the pretrained encoder and new layers; and the random-initialization variant trained the architecture from scratch. The added Tiny components contain about 2.15 million parameters, giving 7.79 million in total. The larger Prithvi-EO-2.0-100M-TL reference contains 86.24 million encoder parameters and 91.97 million in the complete segmentation network. Although its decoder family and width match the Tiny configuration, its input adapters produce a larger non-encoder count, approximately 5.73 million. It was evaluated at 10\% and 100\% labels only, not under Protocol B. Table~\ref{tab:architectures} summarizes these distinctions.

\begin{table*}[!t]
\centering
\caption{Neural architectures and selected learning rates.}
\label{tab:architectures}
{\small\setlength{\tabcolsep}{3.5pt}\renewcommand{\arraystretch}{1.05}
\begin{tabularx}{\textwidth}{@{}lCCCCC@{}}
\toprule
Model & Encoder initialization & Updated components & Total (M) & Trainable (M) & Learning rate \\
\midrule
U-Net & Random & All & 7.76 & 7.76 & $3\times 10^{-4}$ \\
Tiny frozen & Pretrained & Neck + decoder + head & 7.79 & 2.15 & $10^{-3}$ \\
Tiny fine-tuned & Pretrained & All & 7.79 & 7.79 & $10^{-3}/10^{-4}$ \\
Tiny random & Random & All & 7.79 & 7.79 & $3\times 10^{-4}$ \\
Prithvi-100M & Pretrained & All & 91.97 & 91.97 & $10^{-3}/10^{-4}$ \\
\bottomrule
\end{tabularx}}\\[2pt]
\parbox{\textwidth}{\footnotesize For fine-tuned models, the two learning rates refer to new layers / encoder. Tiny encoder: approximately 5.63 M parameters; added components: 2.15 M. Prithvi-100M encoder: 86.24 M; added components: 5.73 M. The decoder family is UPerNet with width 128, but the input adapters and parameter counts are not identical across encoder sizes. Spectral indices have no neural parameters and use zero, training-fitted, or per-chip Otsu thresholds. Prithvi-100M inherits the Tiny learning rate and is evaluated only at 10\% and 100\% labels in Protocol A.}
\end{table*}

\subsection{Training and inference}\label{training-and-inference}

Each optimization step drew 16 random 256$\times$256 crops. Crops with fewer than 5\% valid labeled pixels were redrawn. Spatial augmentation consisted of flips and rotations by multiples of 90 degrees, applied consistently to images and masks. The two-class cross-entropy loss can be written in its equivalent binary form as Equation~\eqref{eq:loss}.

\begin{equation}
\mathcal{L}(\theta)=-\frac{1}{N}\sum_{i\in\Omega}\left[y_i\ln p_i(\theta)+(1-y_i)\ln\left(1-p_i(\theta)\right)\right]
\label{eq:loss}
\end{equation}

In Equation~\eqref{eq:loss}, $\theta$ denotes the trainable model parameters; $\Omega$ is the set of valid labeled pixels in the batch; $N$ is its size; $y_i$ is the reference class of pixel i, with 1 denoting water; and $p_i(\theta)$ is the predicted water probability. Unlabeled and out-of-swath pixels are excluded from $\Omega$. Logarithms in this loss are natural logarithms. No class weights or Dice term were added. This choice provides a straightforward probabilistic objective, but does not itself guarantee calibrated predictions.

Optimization used AdamW \citep{loshchilov2019decoupled}, weight decay 0.05, 100 warm-up steps followed by cosine decay, gradient-norm clipping at 1.0, and mixed precision. The maximum budget was 1,500 steps. Validation was performed every 125 steps; the checkpoint with the greatest validation water IoU was selected. Early stopping required four evaluations without improvement after at least 500 completed steps. These settings define a common practical budget, not a proof that every architecture reached its best possible accuracy.

The base learning rate was selected from $10^{-4}$, $3\times10^{-4}$, and $10^{-3}$ using Protocol A, all training labels, and seed 0. U-Net and randomly initialized Tiny used $3\times10^{-4}$. Frozen and fine-tuned Tiny used $10^{-3}$ for their new layers; the fine-tuned encoder used one-tenth of that rate. Prithvi-100M inherited the fine-tuned Tiny rate rather than receiving an independent search. The study reports 106 runs: 12 selection runs, 48 Protocol A Tiny-scale runs, 40 additional Protocol B runs, and six Prithvi-100M runs. The 94 evaluation runs exclude the selection runs; 30 stopped early.

Full-chip inference used nine overlapping 256-pixel windows with stride 128. Softmax water probabilities were averaged in overlap regions. The decision threshold was fixed at p > 0.5. This same stitching and threshold rule was used for the neural comparisons. The reported implementation retained model configurations, split identities, encoder fingerprints, training histories, metrics, and prediction files in a local Python and cloud-GPU training workflow (Figure~\ref{fig:methodology}C).

\subsection{Accuracy, calibration, and statistical uncertainty}\label{accuracy-calibration-and-statistical-uncertainty}

The primary statistic was water-class IoU obtained from a confusion matrix pooled over valid pixels in each evaluation set. Water F1 was reported alongside precision and recall; the definitions in Equation~\eqref{eq:iou} make their relationship explicit \citep{maxwell2021accuracy2}.

\begin{equation}
\text{IoU}=\frac{\text{TP}}{\text{TP}+\text{FP}+\text{FN}},\qquad F_1=\frac{2\text{TP}}{2\text{TP}+\text{FP}+\text{FN}}
\label{eq:iou}
\end{equation}

TP denotes pixels correctly classified as water, FP denotes non-water pixels classified as water, and FN denotes missed reference-water pixels. Precision is $\text{TP}/(\text{TP}+\text{FP})$, and recall is $\text{TP}/(\text{TP}+\text{FN})$. True negatives contribute to balanced accuracy but not water IoU. A per-chip IoU is undefined when no valid reference pixels exist or when both reference and prediction contain no water; such chips do not contribute to a per-chip average. This convention differs from treating an empty-water chip as perfect. Aggregate and per-chip results answer different questions because their weighting differs.

The distinction between agreement and coverage also arises in the urban-canopy assessment of \citet{narimani2026canopy}, which evaluates candidate delineations separately from the amount of canopy represented. Here, it motivates retaining recall and boundary diagnostics beside IoU rather than allowing a single overlap score to stand for all aspects of a water map.

Water-probability calibration was summarized using 15 equal-width bins and the expected calibration error (ECE) in Equation~\eqref{eq:ece}, alongside reliability diagrams \citep{guo2017calibration}.

\begin{equation}
\text{ECE}=\sum_{b=1}^{B}\frac{n_b}{N}\left|\bar p_b-\bar y_b\right|
\label{eq:ece}
\end{equation}

In Equation~\eqref{eq:ece}, $B$ is the number of bins, $n_b$ is the number of valid pixels in bin $b$, $N$ is the total number of evaluated valid pixels, $\bar p_b$ is their mean predicted water probability, and $\bar y_b$ is their observed water fraction. Empty bins contribute zero. Lower ECE indicates closer average agreement under this binning scheme; it is not a guarantee of accurate boundaries or of reliability within the minority water class. Stored probabilities have resolution 1/255, so very small calibration differences should not be overinterpreted.

Predictive entropy was calculated using Equation~\eqref{eq:entropy}.

\begin{equation}
H(p)=-p\log_2p-(1-p)\log_2(1-p)
\label{eq:entropy}
\end{equation}

Here, $p$ is a pixel's predicted water probability, $H(p)$ is binary entropy in bits, and $\log_2$ is the base-2 logarithm. The endpoint convention is $0\log_2 0=0$. Entropy is zero at probabilities zero and one, and maximal at one-half. It was evaluated as an error-ranking score using the area under the receiver-operating-characteristic curve (AUROC), not interpreted as an error probability. Boundary analysis compared valid pixels within two pixels of a reference water/non-water boundary with interior pixels. The JRC stratum comparison measured recall for reference water marked permanent and for other reference water. These diagnostics used Protocol A, all labels, and seed 0 unless explicitly identified otherwise.

Confidence intervals used 2,000 cluster-bootstrap resamples, reflecting the dependence of pixels within chips \citep{field2007bootstrapping}. For Protocol A, chips were sampled with replacement within each event and their confusion matrices were summed. For multi-seed settings, the statistic was the mean of the seed-specific pooled IoUs on the same resampled chips. These intervals describe test-chip sampling conditional on the sampled events and trained models; the separate seed standard deviations describe training variability. Paired model differences used identical resamples. An interval including zero does not demonstrate equivalence. Under Protocol B, event-specific IoUs were averaged without weighting by chip count, and intervals were obtained by resampling the 11 events. Event-level Wilcoxon tests and chip-property correlations were treated as exploratory. Shared training data between folds and the small number of events limit population-level inference.

\subsection{Computational measurements}\label{computational-measurements}

GPU measurements used a single NVIDIA Tesla T4 with 16 GB memory. Full-chip latency covered a batch of nine inference tiles in mixed precision, with 20 timed repetitions after five warm-up passes. CPU measurements used an Intel Core i9-12900H laptop, 14 threads, and float32 arithmetic, with ten timed repetitions after three warm-up passes. The three Tiny variants shared one CPU architecture timing. Fitted-index CPU measurements included index calculation and thresholding over the 105 test and Bolivia chips. Training time, peak GPU memory, parameter counts, and floating-point operations per 256-pixel tile were also recorded. These are device- and implementation-specific model timings, excluding an operational pipeline's data acquisition, disk access, preprocessing, and export.

\section{Results}\label{results}

\subsection{Accuracy on seen events and the Bolivia holdout}\label{accuracy-on-seen-events-and-the-bolivia-holdout}

U-Net achieved the greatest mean water IoU on the official seen-event test set: 0.856, with a 95\% chip-bootstrap interval of 0.796-0.897 (Table~\ref{tab:seen-event}). Fine-tuned Prithvi-100M reached 0.841, fine-tuned Tiny 0.833, frozen Tiny 0.832, and randomly initialized Tiny 0.809. The best index baseline was fitted MNDWI at 0.787. Precision varied less among the networks than recall: U-Net's recall was 0.909, compared with 0.887 for fine-tuned Tiny. Thus, the U-Net advantage mainly reflected fewer missed water pixels rather than a large precision advantage.

\begin{table*}[!t]
\centering
\caption{Seen-event test performance with all training labels (Protocol A).}
\label{tab:seen-event}
{\small\setlength{\tabcolsep}{3.5pt}\renewcommand{\arraystretch}{1.05}
\begin{tabularx}{\textwidth}{@{}lCCCCCC@{}}
\toprule
Method & Water IoU mean $\pm$ SD & IoU 95\% CI & F1 & Precision & Recall & ECE \\
\midrule
U-Net & 0.856 $\pm$ 0.001 & 0.796--0.897 & 0.922 & 0.936 & 0.909 & 0.014 \\
Tiny frozen & 0.832 $\pm$ 0.004 & 0.772--0.877 & 0.908 & 0.925 & 0.892 & 0.002 \\
Tiny fine-tuned & 0.833 $\pm$ 0.001 & 0.772--0.879 & 0.909 & 0.932 & 0.887 & 0.004 \\
Tiny random & 0.809 $\pm$ 0.003 & 0.748--0.859 & 0.895 & 0.926 & 0.865 & 0.003 \\
Prithvi-100M & 0.841 $\pm$ 0.002 & 0.783--0.886 & 0.914 & 0.936 & 0.892 & 0.005 \\
MNDWI fitted & 0.787 $\pm$ 0.000 & 0.711--0.846 & 0.881 & 0.869 & 0.893 & --- \\
NDWI fitted & 0.773 $\pm$ 0.000 & 0.707--0.824 & 0.872 & 0.864 & 0.88 & --- \\
NDWI zero & 0.735 & 0.635--0.812 & 0.847 & 0.972 & 0.751 & --- \\
MNDWI zero & 0.659 & 0.549--0.749 & 0.795 & 0.675 & 0.965 & --- \\
MNDWI Otsu & 0.386 & 0.281--0.493 & 0.557 & 0.402 & 0.903 & --- \\
NDWI Otsu & 0.308 & 0.220--0.398 & 0.471 & 0.32 & 0.896 & --- \\
\bottomrule
\end{tabularx}}\\[2pt]
\parbox{\textwidth}{\footnotesize 90 test chips from ten events represented in the training pool. Neural metrics are means across three seeds; SD describes seed variation. Intervals are event-stratified chip-bootstrap intervals for the seed-mean pooled IoU, not precision or recall intervals. Fitted indices are deterministic with all labels. ECE uses 15 equal-width water-probability bins; An em dash indicates not applicable or not measured. The evaluation excludes unlabeled and out-of-swath pixels.}
\end{table*}

\begin{figure*}[!t]
\centering
\includegraphics[width=\textwidth]{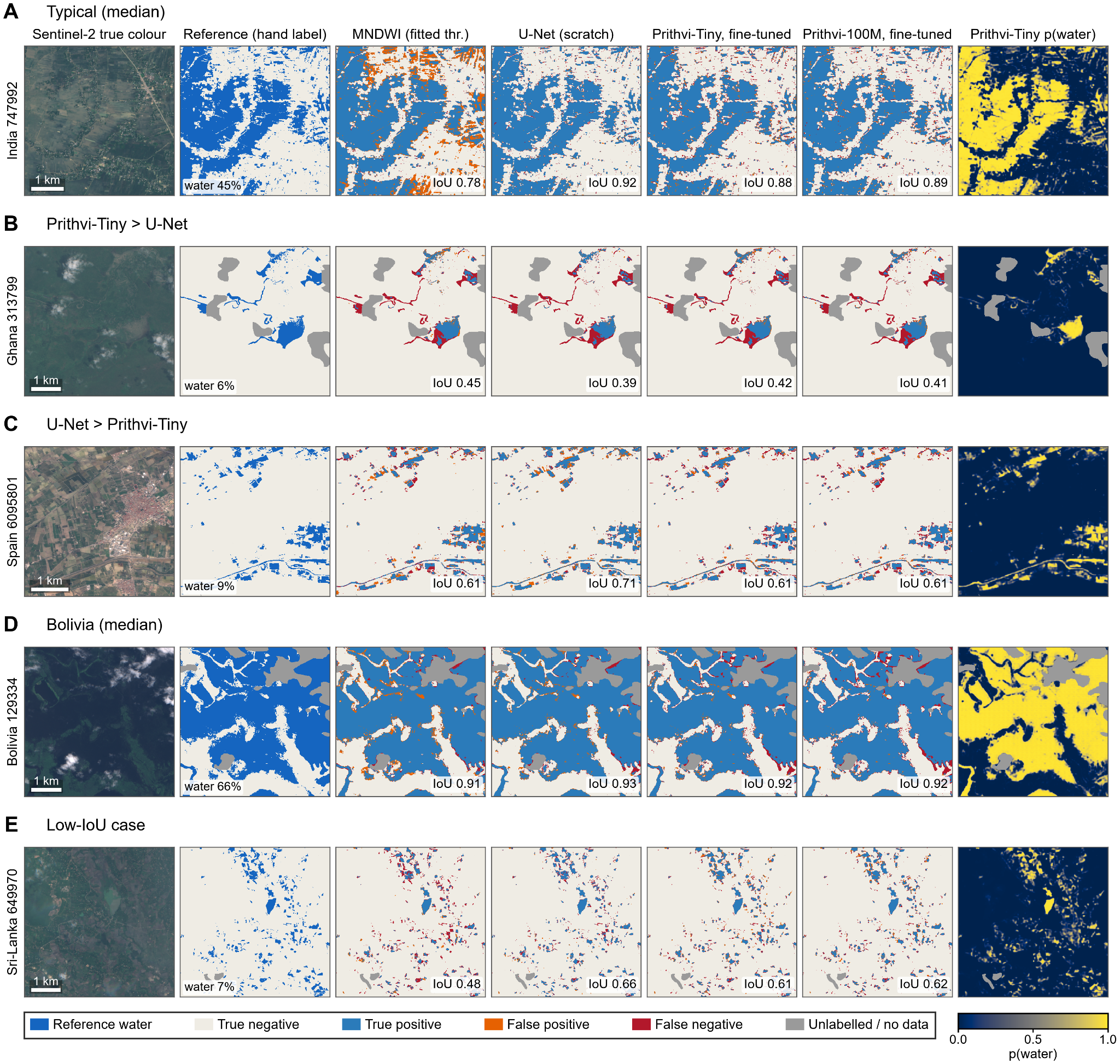}
\caption{Qualitative segmentation comparisons under Protocol A, all labels, seed 0. Columns show true-color imagery, reference masks, MNDWI and neural error maps, and fine-tuned Tiny water probability. Error colors separate true positives, false positives, false negatives, true negatives, and invalid areas. Rows show a median eligible test example (A), the largest eligible relative Tiny and U-Net advantages (B,C), a median eligible Bolivia example (D), and a low-IoU example (E), with at most one chip per event. Selection used at least 5\% reference water and 80\% labeled pixels. These restricted and partly extreme examples do not represent a random sample of all chips. The annotated IoUs are single-chip, seed-0 values, not the three-seed pooled values in Tables~\ref{tab:seen-event} and~\ref{tab:bolivia}.}
\label{fig:segmentation}
\end{figure*}

Figure~\ref{fig:segmentation} places these aggregate results in their spatial context. Broad water bodies are often recovered by several methods, while disagreements occur along margins, narrow channels, and small features. The rows include a median example, relative model advantages, a Bolivia example, and a low-IoU case selected by predefined rules. They are not a random sample of average behavior. Among the 25 eligible seen-event test chips used for this illustration, fine-tuned Tiny exceeded U-Net on one chip, by 0.031 IoU. This is consistent with the aggregate ranking, but the restricted display subset should not replace the full test statistics.

\begin{figure*}[!t]
\centering
\includegraphics[width=\textwidth]{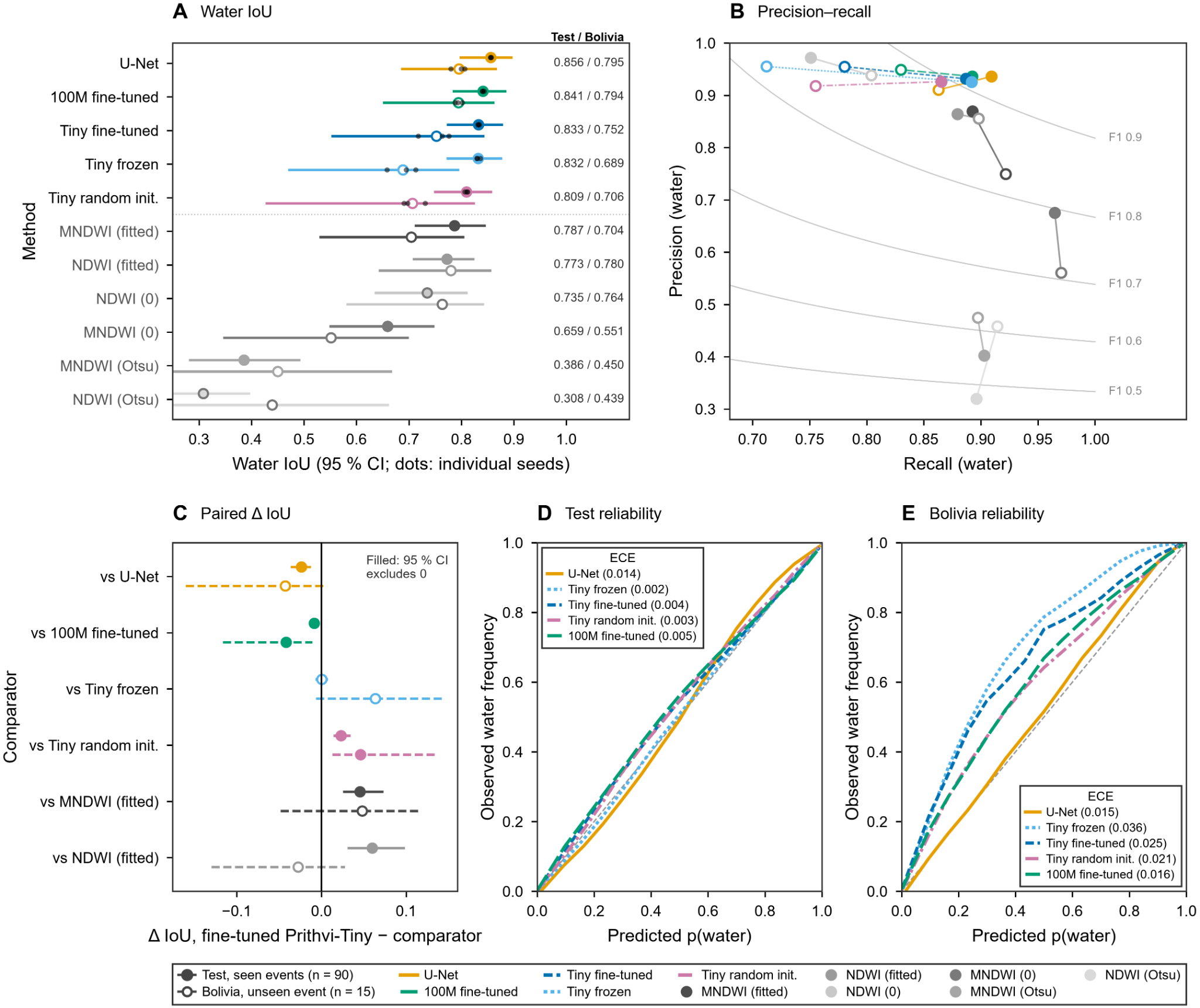}
\caption{Accuracy and calibration with all training labels. (A) Pooled water IoU on 90 seen-event test chips (filled) and 15 Bolivia chips (open), with three-seed means, individual seeds, and chip-bootstrap 95\% intervals. (B) Precision-recall positions and F1 contours. (C) Paired IoU differences between fine-tuned Tiny and each comparator; filled markers identify intervals excluding zero. (D,E) Reliability diagrams for the seen-event test set and Bolivia, pooled across seeds; legends report mean seed ECE. Fifteen equal-width bins are used, and bins with fewer than 1,000 pixels are omitted from the curves. The displayed intervals concern IoU, not precision or recall. Index methods are omitted from probability-calibration panels.}
\label{fig:performance}
\end{figure*}

The paired difference between fine-tuned Tiny and U-Net was -0.024 IoU, with a 95\% interval of -0.037 to -0.012 (Figure~\ref{fig:performance}C). Fine-tuned Tiny exceeded fitted MNDWI by 0.046 (0.026-0.073). Against the identical randomly initialized architecture, the reported gain was 0.023 (0.014-0.034). Paired differences were computed before rounding; they can therefore differ by 0.001 from subtracting the displayed three-decimal means. The comparison supports the pretrained training strategy relative to random initialization under these settings, not superiority over the convolutional baseline.

Bolivia produced a different scale of uncertainty (Table~\ref{tab:bolivia}). U-Net and Prithvi-100M achieved 0.795 and 0.794 IoU, respectively, followed by fitted NDWI at 0.780 and fine-tuned Tiny at 0.752. Frozen Tiny fell to 0.689. The fine-tuned Tiny-U-Net difference was -0.043, but its interval (-0.160 to 0.004) included zero. This does not establish equal performance; it indicates that the 15-chip holdout did not resolve that difference precisely. Pretrained fine-tuned Tiny exceeded its random counterpart by 0.046 (0.013-0.135). A single fitted spectral threshold was competitive in this event even though it was weaker on the seen-event test set.

\begin{table*}[!t]
\centering
\caption{Bolivia event-holdout performance with all training labels (Protocol A).}
\label{tab:bolivia}
{\small\setlength{\tabcolsep}{3.5pt}\renewcommand{\arraystretch}{1.05}
\begin{tabularx}{\textwidth}{@{}lCCCC@{}}
\toprule
Method & Water IoU mean $\pm$ SD & IoU 95\% CI & F1 & ECE \\
\midrule
U-Net & 0.795 $\pm$ 0.014 & 0.685--0.867 & 0.886 & 0.015 \\
Tiny frozen & 0.689 $\pm$ 0.028 & 0.470--0.796 & 0.815 & 0.036 \\
Tiny fine-tuned & 0.752 $\pm$ 0.031 & 0.552--0.844 & 0.859 & 0.025 \\
Tiny random & 0.706 $\pm$ 0.022 & 0.427--0.826 & 0.828 & 0.021 \\
Prithvi-100M & 0.794 $\pm$ 0.007 & 0.651--0.863 & 0.885 & 0.016 \\
MNDWI fitted & 0.704 $\pm$ 0.000 & 0.529--0.806 & 0.827 & --- \\
NDWI fitted & 0.780 $\pm$ 0.000 & 0.643--0.857 & 0.876 & --- \\
NDWI zero & 0.764 & 0.581--0.843 & 0.866 & --- \\
MNDWI zero & 0.551 & 0.346--0.700 & 0.711 & --- \\
MNDWI Otsu & 0.450 & 0.224--0.668 & 0.621 & --- \\
NDWI Otsu & 0.439 & 0.210--0.662 & 0.611 & --- \\
\bottomrule
\end{tabularx}}\\[2pt]
\parbox{\textwidth}{\footnotesize 15 chips from the Bolivia event, absent from gradient fitting, validation, and global learning-rate selection. Neural summaries use three seeds. Confidence intervals describe chip-sampling uncertainty within this event, not uncertainty over all possible future flood events. Model differences with intervals spanning zero are not evidence of equivalence. All metrics use the same validity mask as Table~\ref{tab:seen-event}.}
\end{table*}

\subsection{Sensitivity to the number of fine-tuning chips}\label{sensitivity-to-the-number-of-fine-tuning-chips}

\begin{figure*}[!t]
\centering
\includegraphics[width=\textwidth]{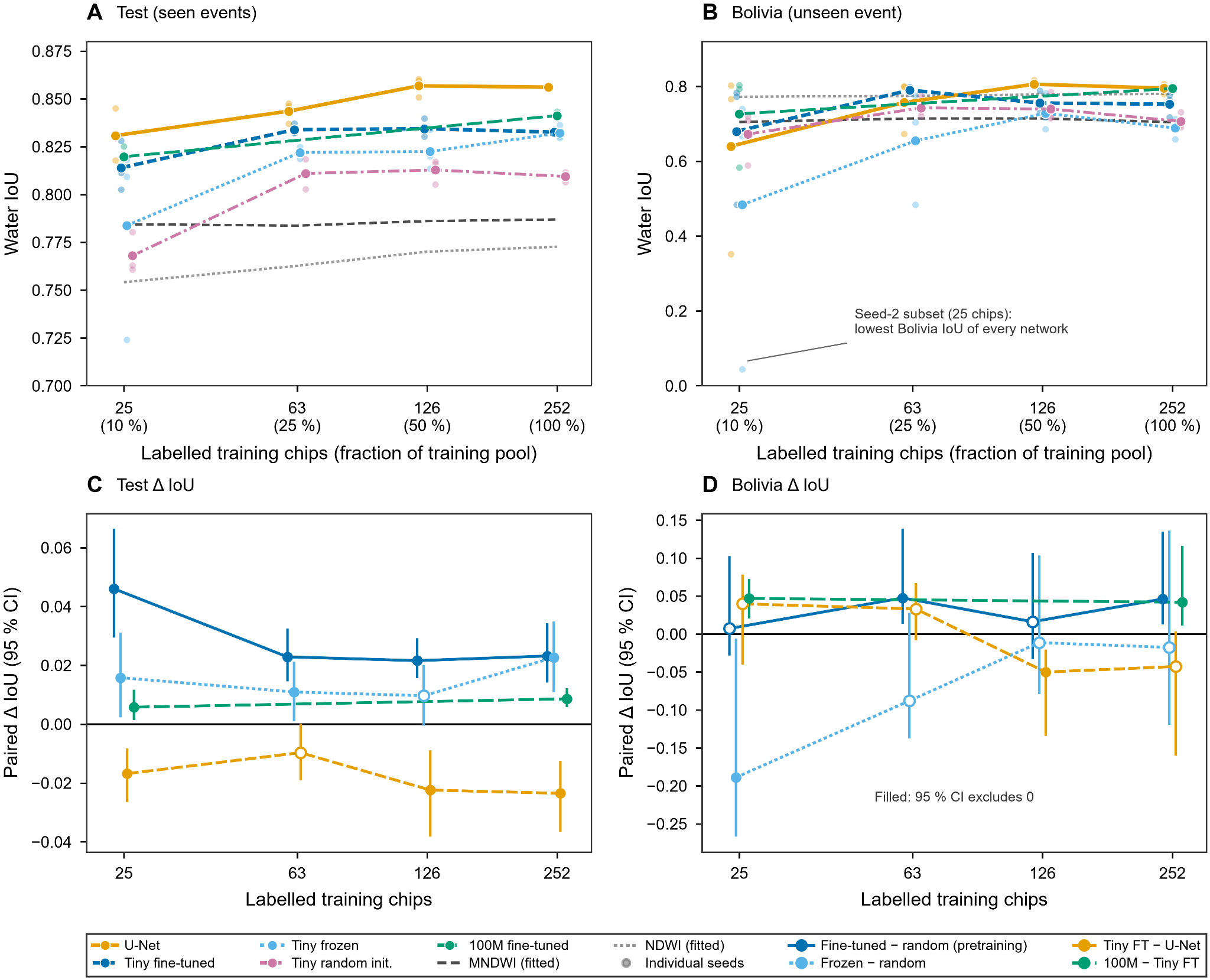}
\caption{Sensitivity to the number of labeled fine-tuning chips. (A,B) Seen-event and Bolivia water IoU for nested training subsets, with three-seed means and individual seed points. The validation pool and global hyperparameter selection remain fixed. Prithvi-100M has measurements only at 25 and 252 chips; its connecting line does not imply intermediate experiments. (C,D) Paired differences for pretrained versus random Tiny, frozen versus random Tiny, fine-tuned Tiny versus U-Net, and 100M versus fine-tuned Tiny. Bars are paired chip-bootstrap 95\% intervals; filled markers indicate intervals excluding zero. Seed variation combines subset and optimization effects. Table~\ref{tab:label-budget} reports means and standard deviations, not confidence intervals for the upper curves.}
\label{fig:label-efficiency}
\end{figure*}

The benefit of pretraining within the Tiny architecture was largest with the smallest labeled subset (Figure~\ref{fig:label-efficiency}; Table~\ref{tab:label-budget}). With 25 training chips, fine-tuned Tiny achieved 0.814 IoU on the seen-event test set, compared with 0.768 for random initialization, a paired gain of 0.046 (0.029-0.066). The reported gain was 0.022-0.023 at larger fractions. Frozen Tiny was less effective with 25 chips, at 0.784, but approached the fine-tuned model with all labels.

U-Net nevertheless retained the highest seen-event mean at every label fraction. Its 25-chip IoU of 0.831 was numerically close to fine-tuned Tiny's 252-chip value of 0.833. This is not an equivalence test, but it is an important practical comparison: the observed pretraining benefit did not establish greater fine-tuning-label efficiency than U-Net. Fitted MNDWI changed little, from 0.784 with 25 chips to 0.787 with 252, suggesting that this transparent spectral rule required relatively few labeled examples in the tested pool.

\begin{table*}[!t]
\centering
\caption{Water IoU under reduced fine-tuning-label budgets (Protocol A).}
\label{tab:label-budget}
{\small\setlength{\tabcolsep}{3.5pt}\renewcommand{\arraystretch}{1.05}
\begin{tabularx}{\textwidth}{@{}lCCCC@{}}
\toprule
\multicolumn{5}{l}{\textbf{A. Seen-event test}}\\
\midrule
Method & 25 chips (~10\%) & 63 chips (25\%) & 126 chips (50\%) & 252 chips (100\%) \\
\midrule
U-Net & 0.831 $\pm$ 0.014 & 0.844 $\pm$ 0.006 & 0.857 $\pm$ 0.005 & 0.856 $\pm$ 0.001 \\
Tiny frozen & 0.784 $\pm$ 0.052 & 0.822 $\pm$ 0.003 & 0.822 $\pm$ 0.010 & 0.832 $\pm$ 0.004 \\
Tiny fine-tuned & 0.814 $\pm$ 0.013 & 0.834 $\pm$ 0.003 & 0.834 $\pm$ 0.005 & 0.833 $\pm$ 0.001 \\
Tiny random & 0.768 $\pm$ 0.011 & 0.811 $\pm$ 0.008 & 0.813 $\pm$ 0.007 & 0.809 $\pm$ 0.003 \\
Prithvi-100M & 0.820 $\pm$ 0.005 & --- & --- & 0.841 $\pm$ 0.002 \\
MNDWI fitted & 0.784 $\pm$ 0.001 & 0.784 $\pm$ 0.003 & 0.786 $\pm$ 0.002 & 0.787 $\pm$ 0.000 \\
NDWI fitted & 0.754 $\pm$ 0.037 & 0.763 $\pm$ 0.014 & 0.770 $\pm$ 0.005 & 0.773 $\pm$ 0.000 \\
\midrule
\multicolumn{5}{l}{\textbf{B. Bolivia holdout}}\\
\midrule
Method & 25 chips (~10\%) & 63 chips (25\%) & 126 chips (50\%) & 252 chips (100\%) \\
\midrule
U-Net & 0.640 $\pm$ 0.250 & 0.757 $\pm$ 0.073 & 0.806 $\pm$ 0.010 & 0.795 $\pm$ 0.014 \\
Tiny frozen & 0.484 $\pm$ 0.382 & 0.655 $\pm$ 0.152 & 0.728 $\pm$ 0.049 & 0.689 $\pm$ 0.028 \\
Tiny fine-tuned & 0.680 $\pm$ 0.170 & 0.790 $\pm$ 0.009 & 0.756 $\pm$ 0.030 & 0.752 $\pm$ 0.031 \\
Tiny random & 0.673 $\pm$ 0.073 & 0.743 $\pm$ 0.029 & 0.739 $\pm$ 0.040 & 0.706 $\pm$ 0.022 \\
Prithvi-100M & 0.727 $\pm$ 0.124 & --- & --- & 0.794 $\pm$ 0.007 \\
MNDWI fitted & 0.705 $\pm$ 0.024 & 0.714 $\pm$ 0.029 & 0.714 $\pm$ 0.017 & 0.704 $\pm$ 0.000 \\
NDWI fitted & 0.772 $\pm$ 0.018 & 0.775 $\pm$ 0.008 & 0.778 $\pm$ 0.003 & 0.780 $\pm$ 0.000 \\
\bottomrule
\end{tabularx}}\\[2pt]
\parbox{\textwidth}{\footnotesize Each cell is mean $\pm$ SD across three seeds. Label fractions change training chips used for gradient updates or index fitting, while the 89-chip validation pool and full-data learning-rate selection remain fixed. Normalization uses all training images without their labels. An em dash indicates not applicable, not measured, or not run. Prithvi-100M was evaluated only at the endpoints. Paired-difference confidence intervals are shown in Figure~\ref{fig:label-efficiency} and discussed in the text; these SDs are not confidence intervals.}
\end{table*}

The Bolivia curves were less stable and not monotonic. With 25 training chips, the seed standard deviation was 0.250 for U-Net, 0.382 for frozen Tiny, and 0.170 for fine-tuned Tiny. Fine-tuned Tiny reached 0.790 at 63 chips but 0.752 at 252. The seed-2, 25-chip runs had the lowest Bolivia IoU for every network; frozen Tiny reached only 0.044 in that run. Because the seed changed both chip selection and optimization, the result demonstrates sensitivity to the combined training realization, not proof that a particular subset alone caused failure. Prithvi-100M was measured only at the two endpoints; the connecting line in Figure~\ref{fig:label-efficiency} does not represent experiments at 25\% or 50\%.

\subsection{Computational trade-offs}\label{computational-trade-offs}

\begin{figure*}[!t]
\centering
\includegraphics[width=\textwidth]{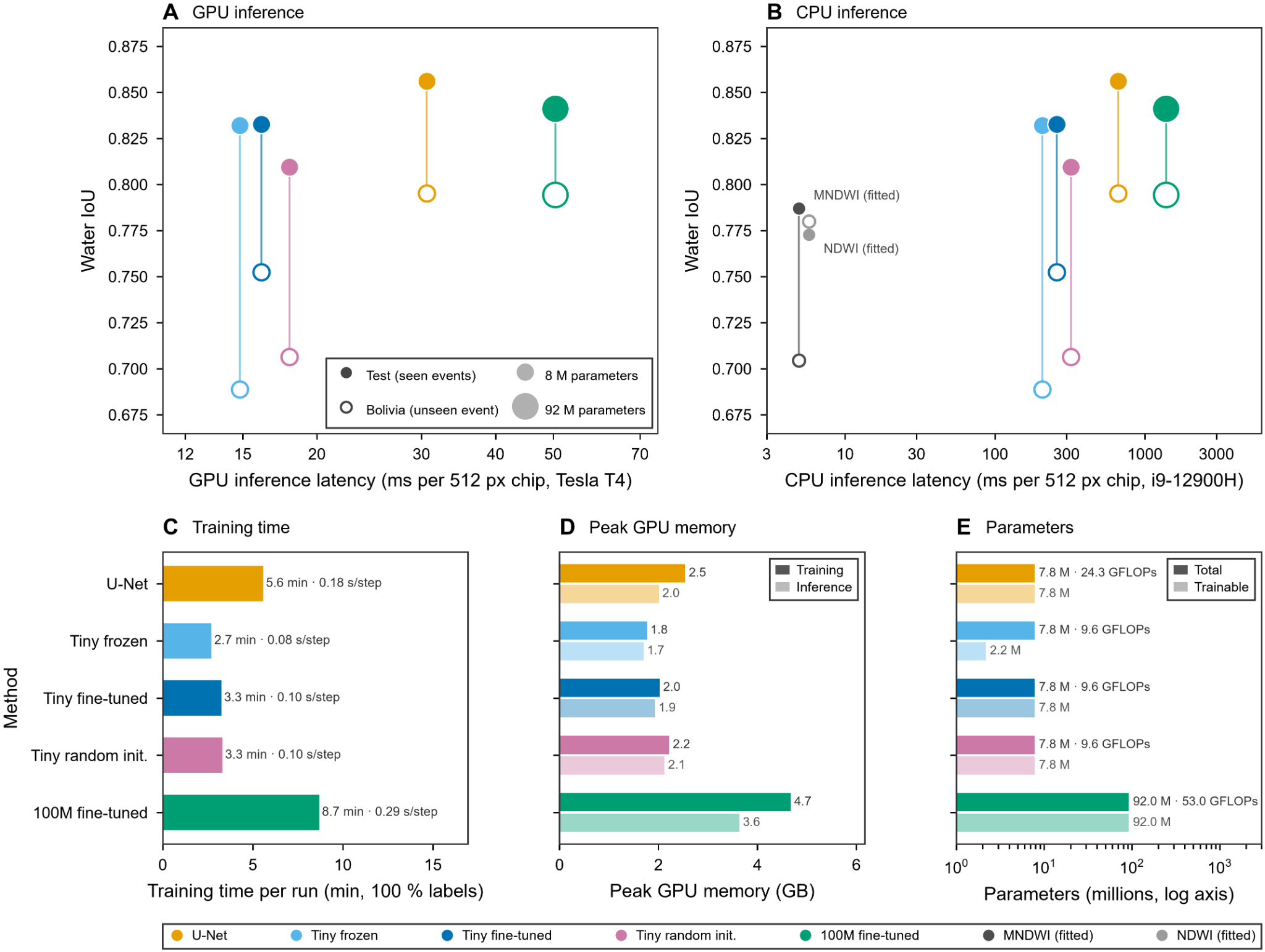}
\caption{Accuracy and computational cost. (A,B) Three-seed pooled IoU against GPU and CPU latency per 512$\times$512 chip. Filled markers denote seen-event test results and open markers Bolivia; marker area encodes parameter count. Tiny variants are offset horizontally for legibility, and share one CPU timing. (C) Median training time with all labels and time per optimization step. (D) Peak GPU memory for training and full-chip inference. (E) Total/trainable parameter counts and per-tile floating-point operations. GPU timings use a Tesla T4, mixed precision, and nine overlapping tiles per chip; CPU timings use an i9-12900H with float32. Measurements exclude data acquisition and file I/O and do not estimate energy consumption.}
\label{fig:efficiency}
\end{figure*}

The main advantage of the compact transformer was computational rather than a higher final overlap score (Figure~\ref{fig:efficiency}; Table~\ref{tab:compute}). Fine-tuned Tiny and U-Net had nearly identical total parameter counts, 7.79 and 7.76 million, but required 9.6 and 24.3 GFLOPs per tile. A full chip took 16.1 ms for fine-tuned Tiny and 30.6 ms for U-Net on the T4, a 1.90-fold latency ratio. On the CPU, 258 versus 665 ms gave a 2.58-fold ratio. The larger Prithvi model required 50.4 ms on the GPU and 1,386 ms on the CPU for a seen-event IoU increase from 0.833 to 0.841.

Frozen Tiny had the shortest median training run, 2.7 min; fine-tuned Tiny required 3.3 min, U-Net 5.6 min, and Prithvi-100M 8.7 min. Peak training memory was 1.8 GB for frozen Tiny, 2.0 GB for fine-tuned Tiny, 2.5 GB for U-Net, and 4.7 GB for Prithvi-100M. Spectral-index baselines remained much faster on the CPU, at 4.9-7.1 ms per chip. Model size alone therefore does not describe the practical trade-off. The tiny transformer was faster than a similarly sized convolutional network in this implementation, but it was neither the most accurate network nor the least expensive method overall.

\begin{table*}[!t]
\centering
\caption{Measured computational cost of the evaluated implementations.}
\label{tab:compute}
{\small\setlength{\tabcolsep}{3.5pt}\renewcommand{\arraystretch}{1.05}
\begin{tabularx}{\textwidth}{@{}lCCCCCC@{}}
\toprule
Method & Parameters total / trainable (M) & GFLOPs / 256-px tile & Training (min) & Training GPU memory (GB) & GPU (ms/chip) & CPU (ms/chip) \\
\midrule
U-Net & 7.8 / 7.8 & 24.3 & 5.6 & 2.5 & 30.6 & 665 \\
Tiny frozen & 7.8 / 2.2 & 9.6 & 2.7 & 1.8 & 15.9 & 258 \\
Tiny fine-tuned & 7.8 / 7.8 & 9.6 & 3.3 & 2.0 & 16.1 & 258 \\
Tiny random & 7.8 / 7.8 & 9.6 & 3.3 & 2.2 & 16.8 & 258 \\
Prithvi-100M & 92.0 / 92.0 & 53.0 & 8.7 & 4.7 & 50.4 & 1386 \\
MNDWI fitted & 0 & --- & --- & --- & --- & 4.9 \\
NDWI fitted & 0 & --- & --- & --- & --- & 5.7 \\
MNDWI Otsu & 0 & --- & --- & --- & --- & 6.7 \\
NDWI Otsu & 0 & --- & --- & --- & --- & 7.1 \\
\bottomrule
\end{tabularx}}\\[2pt]
\parbox{\textwidth}{\footnotesize Training values: median at 100\% labels, including early stopping. GPU: Tesla T4; one 512$\times$512 chip is nine overlapping 256$\times$256 tiles in one mixed-precision batch. CPU: Core i9-12900H, 14 threads, float32; the Tiny variants share one CPU timing. CPU index times include index calculation and thresholding. Timings exclude imagery download, file I/O, and geographic export. M denotes million parameters; an em dash indicates not applicable or not measured. Rounded parameter counts follow these measurements; exact architecture counts are in Table~\ref{tab:architectures}.}
\end{table*}

\subsection{Event-excluded evaluation}\label{event-excluded-evaluation}

Across the 11 Protocol B evaluations, mean event IoU was 0.777 for U-Net, 0.768 for fine-tuned Tiny, 0.765 for frozen Tiny, and 0.725 for randomly initialized Tiny (Table~\ref{tab:protocol-b}; Figure~\ref{fig:generalization}A-C). The paired fine-tuned Tiny-U-Net difference was -0.010 (-0.034 to 0.015), with an exploratory Wilcoxon p-value of 0.32. No event-average advantage of fine-tuned Tiny over U-Net was established. Fine-tuned Tiny did, however, exceed random initialization in all 11 target events, with an average gain of 0.043 (0.032-0.055).

Differences among events were larger than the average difference between these networks. Fine-tuned Tiny exceeded U-Net in Paraguay (0.573 versus 0.522), Somalia (0.722 versus 0.653), and Sri Lanka (0.874 versus 0.864), but fell behind in Pakistan (0.624 versus 0.719) and India (0.731 versus 0.773). In Paraguay, both fitted indices, at 0.609 and 0.615, exceeded all networks. Conversely, fitted NDWI reached only 0.331 in Pakistan. Neither a pretrained encoder nor a simple spectral rule removed event-specific failure.

\begin{table*}[!t]
\centering
\caption{Event-excluded fitting and matched seen/unseen comparisons (Protocol B).}
\label{tab:protocol-b}
{\small\setlength{\tabcolsep}{3.5pt}\renewcommand{\arraystretch}{1.05}
\begin{tabularx}{\textwidth}{@{}lCCCCC@{}}
\toprule
Method & Mean event IoU (95\% CI) & Worst event (IoU) & Events above U-Net & Unseen - seen IoU (95\% CI) & Mean ECE \\
\midrule
U-Net & 0.777 (0.705--0.843) & Paraguay (0.522) & --- & -0.009 (-0.020, -0.001) & 0.030 \\
Tiny frozen & 0.765 (0.705--0.823) & Paraguay (0.564) & 3/11 & -0.011 (-0.018, -0.006) & 0.010 \\
Tiny fine-tuned & 0.768 (0.703--0.828) & Paraguay (0.573) & 3/11 & -0.004 (-0.014, +0.006) & 0.009 \\
Tiny random & 0.725 (0.657--0.792) & Paraguay (0.519) & 1/11 & -0.008 (-0.028, +0.010) & 0.013 \\
Prithvi-100M & not run & not run & not run & not run & not run \\
MNDWI fitted & 0.730 (0.671--0.792) & Pakistan (0.602) & 2/11 & -0.002 (-0.008, +0.003) & --- \\
NDWI fitted & 0.695 (0.595--0.783) & Pakistan (0.331) & 1/11 & -0.014 (-0.034, +0.001) & --- \\
\bottomrule
\end{tabularx}}\\[2pt]
\parbox{\textwidth}{\footnotesize Event means give equal weight to the 11 target events; CIs resample events. Each target uses seed 0 and the fixed-pool exclusions described in Section 2.3. Global learning-rate selection was reused from Protocol A; this is not fully nested event-blind tuning. Unseen - seen compares identical official test chips under A seed 0 and the corresponding B model, with a paired chip bootstrap. Prithvi-100M was not run under B. An em dash indicates not applicable or not measured.}
\end{table*}

A paired comparison on the same 90 official test chips separated changes in fitting from changes in the evaluated images. Using seed 0, U-Net decreased from 0.856 to 0.847 when the respective event was excluded; fine-tuned Tiny decreased from 0.834 to 0.830. These values differ from the three-seed summaries by design. The small decreases should not be read as proof of broad geographic robustness: the global learning-rate selection was shared across protocols, and the full held-out event sets still showed substantial variation.

\subsection{Reliability, water boundaries, and difficult cases}\label{reliability-water-boundaries-and-difficult-cases}

On the seen-event test set, mean ECE ranged from 0.002 for frozen Tiny to 0.014 for U-Net (Figure~\ref{fig:performance}D). On Bolivia, it increased to 0.025 for fine-tuned Tiny and 0.036 for frozen Tiny, compared with 0.015 for U-Net. The reliability curves above the diagonal indicate that the Tiny models tended to assign water probabilities below the observed water frequency in those bins. Across Protocol B events, however, mean ECE was 0.009 for fine-tuned Tiny and 0.030 for U-Net. This change in ranking shows why calibration in one holdout is not a universal property of an architecture.

\begin{figure*}[!t]
\centering
\includegraphics[width=\textwidth]{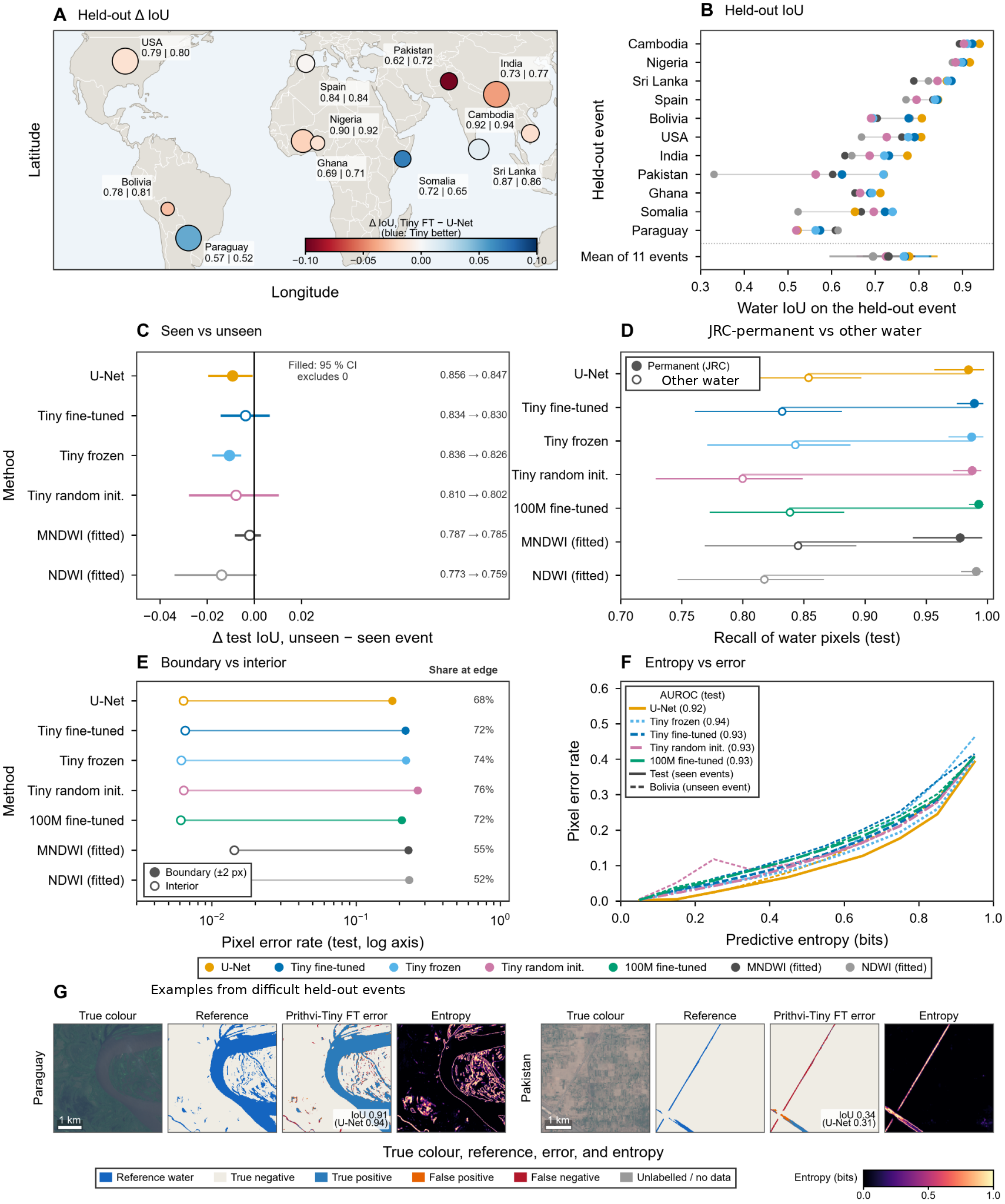}
\caption{Event-level performance and reliability. (A,B) Protocol B event IoUs and paired Tiny-U-Net differences; the event mean gives equal weight to all 11 targets. These are event-excluded fits with shared Protocol A learning-rate selection, not fully nested tuning. (C) Changes on identical official test chips under seen-event versus event-excluded seed-0 fitting. (D) Recall for JRC-permanent and other reference water; the latter is not verified newly inundated land. (E) Boundary and interior error rates and the fraction of errors near boundaries. (F) Pixel error versus entropy, with error-detection AUROC. Panels D-F use Protocol A, all labels, seed 0. (G) Clear-image examples from Paraguay and Pakistan, two low-performing events; the Paraguay example itself has high IoU.}
\label{fig:generalization}
\end{figure*}

Pixels within two pixels of a water boundary accounted for 7.1\% of valid test pixels but 68-76\% of neural-model errors (Figure~\ref{fig:generalization}E). Fine-tuned Tiny's boundary error rate was 22.0\%, compared with 17.8\% for U-Net, whereas their interior rates were similar, 0.65\% and 0.63\%. Extent and boundary behavior are therefore separate aspects of map quality, a distinction also made in the visible-boundary mapping work of \citet{narimani2026farmland}. Here, the flood-specific evidence is the contrast between boundary and interior errors in Figure~\ref{fig:generalization}E.

JRC-permanent reference water was easier to recover than other reference water (Figure~\ref{fig:generalization}D). Neural recall was 0.985-0.993 for the permanent stratum and 0.800-0.854 for its complement. The difference cannot be interpreted as a measured penalty for newly flooded land without an independent pre-event reference. Per-chip IoU was positively associated with reference-water fraction (Spearman $\rho$ 0.62-0.73), while the bright-blue reflectance proxy and unlabeled fraction showed no clear association (absolute $\rho$ at most 0.14; p at least 0.21). These exploratory associations do not establish turbidity, clouds, or vegetation as the causes of particular errors.

Entropy ranked erroneous pixels usefully, with AUROC 0.92-0.94 on seen-event test chips and 0.91-0.95 on Bolivia (Figure~\ref{fig:generalization}F). For fine-tuned Tiny, pixels at or above 0.5 bits represented 5.5\% of the test pixels but contained 71\% of its errors. Their error rate was 28\%, compared with 0.7\% elsewhere. Nevertheless, low entropy did not exclude failure. The Paraguay example in Figure~\ref{fig:generalization}G had high IoU (0.91), despite belonging to a difficult event, whereas the Pakistan canal example had IoU 0.34. These are examples from low-performing events, not necessarily their worst individual chips.

\section{Discussion}\label{discussion}

\subsection{What pretraining did, and did not, provide}\label{what-pretraining-did-and-did-not-provide}

The results support pretraining within the compact transformer, but not a general advantage over U-Net. Fine-tuned Tiny exceeded the identical random-initialization architecture on the seen-event test set at every label fraction and in every Protocol B target event. The largest gain occurred with 25 training chips. Yet U-Net achieved higher seen-event mean IoU at every fraction and a slightly higher event-average value. This distinction is consistent with benchmarks showing that a large pretraining corpus does not automatically displace a strong supervised baseline \citep{marsocci2026pangaea,kaushik2026assessing}. The useful finding is not that pretraining fails, but that its benefit depends on the comparator.

The random-initialization control strengthens this interpretation, with an important qualification. Architecture, decoder family, and maximum update budget were controlled, but pretrained and random models used different selected learning rates and encoder-rate policies. The result therefore estimates the benefit of the pretrained adaptation strategy under the tested tuning procedure, rather than an experiment changing only the initial weights. A fully factorial optimization study would be needed to separate these effects. Similarly, the larger model's inherited learning rate and restricted evaluation do not establish the maximum attainable performance of Prithvi-100M.

Results from other spectral tasks should not be used to fill that evidential gap. The multi-trait grapevine study of \citet{farajpoor2025multitrait} concerns relationships between leaf spectra and physiological or nutrient targets, rather than water segmentation. This emphasizes a target-specific interpretation of spectral representations: success with physiological traits does not imply the same architecture ranking for open-water boundaries.

Boundary behavior offers a plausible explanation for the observed accuracy gap. At the supplied resolution, one 16-pixel transformer patch covers roughly 160 m in the north-south direction, while the decoder must recover much finer edges. U-Net carries spatial detail through skip connections. Our similar interior error rates but different boundary rates are consistent with a local-detail limitation, although they do not prove that patch tokenization alone caused it. Hybrid approaches such as U-Prithvi explicitly combine pretrained representations with convolutional detail \citep{kostejn2025uprithvi}. Such a model would be a logical next comparison, not an untested explanation to present as a result.

\subsection{Label savings must be defined against the actual information budget}\label{label-savings-must-be-defined-against-the-actual-information-budget}

The smallest-subset experiments show that relatively few labeled training chips can support useful water segmentation in this benchmark. They do not show that the whole study could have been developed with only 25 labeled chips. The validation set remained labeled, learning rates were selected with full-label Protocol A training, and normalization used the full image pool. This is a fine-tuning-label experiment under shared development information. Future annotation-budget studies should vary validation and tuning resources as well as gradient-training labels.

The Bolivia variability also cautions against drawing a strong conclusion from a single subset. Some reduced-label means exceeded the full-label result, while their seed variability differed greatly. Because subset selection and optimization seeds were coupled, both can contribute. The broader small-data remote-sensing literature emphasizes that experimental design, sample diversity, and transfer choices are as important as selecting a nominally data-efficient architecture \citep{safonova2023ten}. Here, no method was reliably more transferable simply because its seen-event score changed little with fewer chips.

\subsection{Geographic transfer is a property of the evaluation as well as the model}\label{geographic-transfer-is-a-property-of-the-evaluation-as-well-as-the-model}

The observed event spread, including cases where indices outperformed every network, argues for reporting event-specific results beside a pooled score. This agrees with cross-dataset flood-model assessments and with the wider caution against extrapolating environmental maps beyond their evaluated domain \citep{portalesjulia2025understanding,meyer2022machine}. An unweighted event mean answers a different question from a pixel-pooled IoU: it gives small events equal weight rather than allowing water-rich or larger collections to dominate.

Protocol B provides useful evidence about excluding an event from fitting, but its shared learning-rate selection prevents a claim of completely unseen event-level model development. Its non-Bolivia folds also omit Bolivia from the training pool, as dictated by the fixed benchmark manifests. These design details are not cosmetic. They distinguish the reported evaluation from fully nested leave-one-event-out validation. Replication should preserve them for comparability, while a future study should select hyperparameters inside each event-exclusion fold and repeat each fold with several seeds.

The target itself also constrains transfer claims. Work linking wildfire severity to snow persistence distinguishes predicting an environmental state from attributing a disturbance-related response \citep{farajpoor2026creekfire}. The parallel in this study is the difference between recognizing surface water and establishing that a flood newly created it. Event-level generalization of the former does not validate the latter.

\subsection{A practical efficiency advantage, not an operational guarantee}\label{a-practical-efficiency-advantage-not-an-operational-guarantee}

Under the recorded implementation, fine-tuned Tiny reduced GPU latency by about 47\% and CPU latency by about 61\% relative to U-Net, while losing a reported 0.024 seen-event IoU. The decision between them is therefore application-specific. A resource-limited analyst may prefer the compact transformer when that accuracy difference is acceptable, whereas a user prioritizing the strongest seen-event overlap may prefer U-Net. Fitted spectral indices remain useful low-cost reference methods and, in some held-out events, competitive predictors.

These measurements do not include the cost of obtaining cloud-free observations, transferring imagery, reconstructing full scenes, or validating an operational map. They also do not measure electrical energy or carbon emissions. A shorter inference time should not be translated into an energy-reduction percentage without power measurements. GPU and CPU ratios are specific to the tested hardware, precision, batching, and software. The study supports a compute-accuracy comparison for these implementations, not a general claim of operational readiness or reduced disaster losses.

\subsection{Reliability maps support review but do not replace reference evidence}\label{reliability-maps-support-review-but-do-not-replace-reference-evidence}

Entropy concentrated many errors into a relatively small subset of pixels. This gives it a practical use: prioritizing inspection of uncertain margins rather than treating every mapped pixel as equally trustworthy. However, AUROC measures ranking, not whether an entropy threshold provides a reliable probability of error. Low ECE can also coexist with locally important mistakes, particularly when non-water pixels dominate. Calibration under shift is known to depend on the target distribution \citep{ovadia2019can}, and the differing Bolivia and Protocol B rankings reinforce the need to inspect it locally.

The distinction between a mapped association and an explanation of a hazard is similarly important in multisource assessments of wildfire impacts and exposure \citep{farajpoor2026kincade}. In this study, lower recall outside the JRC-permanent-water stratum is an observed association. It cannot, by itself, identify water depth, turbidity, land cover, or flood origin as the mechanism. Those variables would require independent measurements.

The hand-labeled masks are a reference, not error-free physical truth. Boundary ambiguity, unobserved water beneath vegetation, and cloud obstruction affect both interpretation and evaluation \citep{shastry2023mapping}. The bootstrap intervals account for sampled chips or events, not annotation uncertainty, and the stored probabilities' quantization limits very fine calibration comparisons. A clear-sky benchmark can establish agreement with visible reference water without demonstrating complete inundation mapping during cloudy flood peaks.

\subsection{Limits and next tests}\label{limits-and-next-tests}

Several limits define the scope of the findings. The study contains only 11 events, unevenly represented by chips. Protocol A has three seeds, Protocol B one, and all models use a restricted update budget. Fine-tuned and random models do not share identical optimization policies. The 100M model lacks a separate learning-rate search and event-exclusion experiment. The supplied HLS-pretrained encoder receives Level-1C images at a different spatial scale; domain adaptation literature explains why normalization alone need not remove this mismatch \citep{tuia2016domain}. A matched HLS-input experiment would be needed to estimate its effect.

Single-date input is another deliberate restriction. Satellite time-series methods can represent temporal development, as illustrated in the branched-broomrape study of \citet{narimani2025broomrape}, but that is a different use of temporal information from the present experiment. For floods, the relevant next test would use pre-event and event-date observations to separate persistent water from change, rather than infer a temporal capability from a model name.

The most informative extensions are therefore targeted: nested event-level model selection, repeated event-exclusion seeds, an independently verified pre-event water reference, and optical-radar or multitemporal inputs evaluated under the same masks and metrics. These tests would address identified uncertainties without obscuring the current result with a more elaborate architecture. The present evidence is sufficient to describe a useful computational trade-off, but not to declare a universally best flood-mapping model.

\section{Conclusions}\label{conclusions}

Compact Prithvi pretraining improved surface-water segmentation relative to random initialization of the same transformer, especially with fewer fine-tuning chips. It did not consistently outperform a similarly sized U-Net: U-Net led on the seen-event test set, while the event-average difference was unresolved by the reported interval. The principal advantage of fine-tuned Tiny was lower measured computational demand, with approximately half the GPU latency and two-fifths the CPU latency of U-Net under the tested implementation.

Event-specific errors, boundary behavior, and calibration are essential to interpreting that trade-off. Spectral indices remained valuable references and sometimes outperformed all networks on a difficult event. Entropy helped identify locations needing review, but did not establish correctness. Finally, these experiments map reference surface water during floods, not independently verified new inundation. Shared hyperparameter selection and fixed validation resources further limit claims about wholly unseen-event performance and total annotation savings. A defensible model choice should therefore combine accuracy, computation, and the similarity of the intended setting to the evaluated events, rather than rely on the presence or size of a pretrained encoder alone.

\FloatBarrier

\section*{Data availability statement}\label{data-availability-statement}
The study uses the public Sen1Floods11 version 1.1 dataset \citep{bonafilia2020sen1floods11}, the public NASA--IBM Prithvi-EO-2.0 model releases \citep{szwarcman2024prithvi}, and the JRC-derived permanent-water layer distributed with Sen1Floods11 \citep{pekel2016highresolution}. Copernicus elevation data and Natural Earth geographic layers provide cartographic context in Figure~\ref{fig:study-area}. Derived numerical tables, split manifests, statistical outputs, an archived copy of the analysis code, and a sample of derived chip previews (true-colour excerpts, label visualisations, predicted masks, and georeferenced probability/error layers) are openly available on Zenodo at \url{https://doi.org/10.5281/zenodo.23653544}. The original Sen1Floods11 GeoTIFF collection remains available from its public Google Cloud bucket. Redistribution of source imagery, pretrained weights, and geographic layers follows the terms of the original providers.

\section*{Code availability statement}\label{code-availability-statement}
Python routines for water-index baselines, segmentation metrics, event-stratified splits, and the U-Net / Prithvi model factory used in this study are available at \url{https://github.com/MohammadrezaNarimaniUCDavis/Prithvi_Tiny_Floodwater_GeoAI}. An archived replication snapshot is included in the Zenodo record \url{https://doi.org/10.5281/zenodo.23653544}.

\section*{Author contributions}\label{author-contributions}
Parastoo Farajpoor: Conceptualization, Methodology, Validation, Visualization, Writing---original draft, Writing---review and editing.\\
Mohammadreza Narimani: Conceptualization, Data curation, Formal analysis, Methodology, Software, Validation, Visualization, Writing---original draft, Writing---review and editing, Project administration.

\section*{Funding}\label{funding}
The authors declare that no external financial support was received for the research, authorship, or publication of this article.

\section*{Conflict of interest}\label{conflict-of-interest}
The authors declare that the research was conducted in the absence of commercial or financial relationships that could be construed as a potential conflict of interest.

\section*{Ethics statement}\label{ethics-statement}
The study used geospatial and environmental observations and did not involve human participants, identifiable personal information, or experimental work with animals.

\section*{Generative AI statement}\label{generative-ai-statement}
During the preparation of this work, the authors used ChatGPT to improve grammatical accuracy, refine sentence structure, and enhance visualizations. All AI-generated revisions were thoroughly reviewed and edited by the authors to ensure relevance and accuracy.

\section*{Acknowledgments}\label{acknowledgments}
The authors acknowledge the creators of Sen1Floods11, the NASA--IBM Prithvi team, and the developers of TerraTorch and the open-source geospatial software used in the study. Figure~\ref{fig:study-area} contains modified Copernicus DEM data and Natural Earth geographic data. The study contains modified Copernicus Sentinel data. Terrain products were produced using Copernicus WorldDEM-30, \textcopyright\ DLR e.V.\ 2010--2014 and \textcopyright\ Airbus Defence and Space GmbH 2014--2018, provided under Copernicus by the European Union and ESA.

\section*{Abbreviations}\label{abbreviations}
AUROC, area under the receiver-operating-characteristic curve; ECE, expected calibration error; GFM, geospatial foundation model; HLS, Harmonized Landsat Sentinel-2; IoU, intersection-over-union; JRC, Joint Research Centre; LOEO, leave-one-event-out; MNDWI, modified normalized difference water index; NDWI, normalized difference water index; TOA, top-of-atmosphere; UPerNet, Unified Perceptual Parsing Network; ViT, vision transformer.

\bibliographystyle{IEEEtranN}
\bibliography{references}

\end{document}